\documentclass[9pt,conference]{IEEEtran}
\IEEEoverridecommandlockouts
\usepackage{cite}
\usepackage{amsmath,amssymb,amsfonts}
\usepackage{algorithmic}
\usepackage{graphicx}
\usepackage{textcomp}

\usepackage{booktabs}
\usepackage{color}
\usepackage{url}
\usepackage{multirow}
\usepackage{diagbox}
\usepackage{float}
\usepackage{subfig}
\usepackage{enumitem}

\usepackage[numbers,square]{natbib}

\usepackage[table]{xcolor} 
\definecolor{mycolor}{RGB}{230, 230, 255} 

\def\BibTeX{{\rm B\kern-.05em{\sc i\kern-.025em b}\kern-.08em
    T\kern-.1667em\lower.7ex\hbox{E}\kern-.125emX}}
\begin{document}

\title{RF-GML: Reference-Free Generative Machine Listener\\

\thanks{*The author worked at Dolby Germany GmbH during the development of this technology and is now at Fraunhofer IIS in Erlangen, Germany.}
}

\author{\IEEEauthorblockN{Arijit Biswas and Guanxin Jiang*}
\IEEEauthorblockA{\textit{Dolby Germany GmbH, Nürnberg, Germany} \\
arijit.biswas@dolby.com
}



}

\maketitle

\begin{abstract} 
This paper introduces a novel reference-free (RF) audio quality metric called the RF-Generative Machine Listener (RF-GML), designed to evaluate coded mono, stereo, and binaural audio at a 48 kHz sample rate. RF-GML leverages transfer learning from a state-of-the-art full-reference (FR) Generative Machine Listener (GML) with minimal architectural modifications. The term ``generative'' refers to the model's ability to generate an arbitrary number of simulated listening scores. Unlike existing RF models, RF-GML accurately predicts subjective quality scores across diverse content types and codecs. Extensive evaluations demonstrate its superiority in rating unencoded audio and distinguishing different levels of coding artifacts. RF-GML’s performance and versatility make it a valuable tool for coded audio quality assessment and monitoring in various applications, all without the need for a reference signal.
\end{abstract}


\section{Introduction}
\label{section:Introduction}

Algorithms for objective quality assessment are classified as either \textit{full-reference} (aka intrusive) or \textit{reference-free} (aka non-intrusive). Full-reference (FR) algorithms~\cite{peaq},\cite{ViSQOLAudioIEEE},\cite{polqa} require both a reference and a test signal, while reference-free (RF) algorithms~\cite{non-intru_ms},\cite{Fu2018QualityNetAE},\cite{Wawenets},\cite{NISQA},\cite{SESQA} rely solely on the test signal. FR methods typically outperform RF methods, especially in evaluating lossy coding, where the absence of a reference signal makes it difficult to distinguish between coding artifacts and intentional production choices. However, RF metrics are essential for quality monitoring in scenarios like internet streaming. They could also add value in applications such as large-scale music distribution and archiving, provided they effectively identify high-quality unencoded audio over coded audio.

As noted in~\cite{noresqa}, RF models struggle to learn the implicit references that subjects use (based on prior listening experience) when assessing quality. To address this, metrics like NORESQA~\cite{noresqa} and NOMAD~\cite{nomad} utilize \textit{non-matching references}, which are any clean reference speech signals. NOMAD has been shown to outperform other non-matching reference methods~\cite{noresqa} in speech quality assessment. However, its main drawback is that predictions depend on the non-matching reference signal, requiring the model to use multiple (e.g., 900) non-matching references~\cite{nomad}, which can increase computational complexity. This motivated us to revisit RF metrics, focusing on evaluating coded audio across all content types at a 48 kHz sample rate.

While most RF models are designed for speech, typically at a 16 kHz sample rate, and focus on estimating quality degraded by noise or enhanced by deep noise suppression methods, some models address a broader range of impairments. Specifically for evaluating coded speech,~\cite{no-ref-pesq-dnn} introduced an RF model to estimate PESQ (an FR metric) scores from various wideband codecs at a 16 kHz sample rate. WaweNet~\cite{Wawenets} and NISQA~\cite{NISQA} can also predict the quality of coded speech, with NISQA predicting transmitted speech quality at a 48 kHz sample rate. Recently, SESQA~\cite{SESQA}, an RF metric for speech at a 48 kHz sample rate, was proposed, improving accuracy over NISQA and other RF metrics by training on multiple objectives, such as Mean Opinion Score (MOS), ranking, and FR metrics.

For evaluating coded audio at a 48 kHz sample rate, widely used objective audio quality assessment algorithms like PEAQ~\cite{peaq} and ViSQOL~\cite{ViSQOLAudioIEEE} are intrusive. Recently, full-reference models with inception-based~\cite{Incept2} convolutional architectures adapted for audio~\cite{InSE-NET},\cite{StereoInSE-NET},\cite{gml} have significantly outperformed both PEAQ and ViSQOL. To the best of the authors’ knowledge, only two prior studies have explored RF methods for assessing coded audio, including music at a 48 kHz sample rate. In~\cite{Mumtaz}, the authors developed a mono RF model for user-generated content using a dataset with subjective quality scores collected online. While this research included codec distortion, it lacked details about the codecs used, and, as noted by the authors (Fig. 8 in~\cite{Mumtaz}), the subjective quality of coded audio did not scale with bitrates, raising concerns about the suitability of the data for training and testing an RF metric for codecs. In~\cite{no-ref-audio-lossy}, the authors proposed a method for assessing mono-coding artifacts specifically for music signals. For training, they used music signals processed with various codecs (and bitrates) and low-pass filtered versions, with ViSQOL-v3~\cite{ViSQOLgithub} scores as quality labels. However, since the model was trained to mimic ViSQOL, which outputs a maximum score of 4.73 (not the highest MOS of 5.0)~\cite{InSE-NET}, saturation in predicted quality was expected and observed. Additionally, the reported results were based on averaging outputs from 15 models, indicating high complexity. Consequently, we aim to develop an RF model trained on individual MUSHRA (Multiple Stimuli with Hidden Reference and Anchor)~\cite{MUSHRA} listening test scores of coded stereo and binaural signals collected in a controlled lab setting, and we present the results of these models individually.

Our contributions to this paper are as follows:
\begin{itemize}
\item We introduce RF-GML, a novel RF audio quality metric transfer-learned from a state-of-the-art FR Generative Machine Listener (GML)~\cite{gml} with minimal architectural changes. Like GML, RF-GML is trained on individual subjective scores and can predict a distribution of scores.
\item RF-GML evaluates coded mono, stereo, and binaural signals at a 48 kHz sample rate across all content types, a capability not previously proposed for an RF model.
\item We report prediction accuracy against internal MUSHRA tests and comprehensive MUSHRA tests conducted by MPEG for popular audio codecs~\cite{usac_lt}. Additionally, we benchmark our model against a state-of-the-art RF model for speech at a 48 kHz sample rate, evaluating performance for both coded speech and general audio.
\item We propose including the mean quality of unencoded audio as a performance metric alongside the usual correlation metrics. We hope our approach to testing RF models offers new insights and inspires further research.
\end{itemize}

The paper is organized as follows: Section~\ref{section:RF-GML} introduces the generative modeling approach for predicting audio quality with RF-GML. Section~\ref{section:Datasets} describes the data used for RF-GML model training and evaluation. Experimental results and analysis are presented in Section~\ref{section:Results}, and the conclusion is provided in Section~\ref{section:ConclusionDiscussion}.

\section{RF-GML}
\label{section:RF-GML}

In the subjective audio quality evaluation of codecs, both mean quality scores and confidence intervals (CIs) are informative, but most predictors only estimate the mean. For RF speech quality modeling at a 16 kHz sample rate,~\cite{faridee22_MOSdistr} used histogram matching to model the distribution of subjective scores. In GML~\cite{gml}, a generative modeling method is used to derive both the mean score and CIs. Unlike~\cite{faridee22_MOSdistr}, GML’s task is to efficiently simulate listener scores for a given reference-coded signal pair, going beyond simple regression on mean scores by leveraging maximum likelihood training, which adjusts to variability in the number of listeners.

The GML concept can be easily adapted to RF-GML. For a test signal $y$, the RF-GML model provides a probability distribution of scores $s$ using a probability density function parameterized by model parameters $\theta$: 
\begin{equation} 
p_\theta(s \vert y). 
\end{equation} 
The model’s generative nature allows for simulating a listening test with $N$ listeners by sampling the model $N$ times. Alternatively, using an explicit output distribution enables direct derivation of desired statistics from the model's output parameters. We use a two-parameter logistic distribution for density modeling, as it has been shown to perform better than the Gaussian~\cite{gml}. Like GML, the RF-GML model outputs a mean value $\mu$ and the logarithm of the logistic scale parameter $a$. Model parameters $\theta$ are trained using negative log-likelihood (NLL) loss, with each pair $(y,s)$ contributing according to \eqref{losslogistic}. 
\begin{align} 
\label{losslogistic} 
L_\text{logistic} &= \log (4a) + 2 \log \operatorname{sech} \biggl( \frac{s-\mu}{2a} \biggr). 
\end{align} 
To predict mean quality, we use the model output $\mu$. A 95\% CI can be calculated using a t-distribution, based on the number of listeners $N$ and the model’s standard deviation ($\pi a/\sqrt{3}$ for a logistic distribution).

Adapting GML to RF-GML involves three key steps:
\begin{itemize} 
\item Remove the reference signal channels from the input layer and retain only the degraded input signal channels. 
\item Transfer the learned weights from GML to RF-GML for initialization, leveraging the domain knowledge acquired in GML. Specifically, we propose transferring the weights of the degraded input channels from the first block in GML rather than training the model from scratch. 
\item Retrain RF-GML using \eqref{losslogistic}. 
\end{itemize}

\section{Datasets}
\label{section:Datasets}

\subsection{Training set}
\label{subsection:trainingset}

We conducted our internal MUSHRA listening tests at a 48 kHz sample rate using headphones. Each test featured an unencoded hidden reference, 3.5 kHz and 7 kHz low-pass filtered versions of the reference, and one or more coded signals. The stereo codecs in the training set included AAC~\cite{AAC}, HE-AAC v1 and v2~\cite{HE-AAC}, and Dolby AC-4~\cite{ac4_IEEE}, covering various bitrates and all content types.

Binaural tests involved object-based immersive Dolby Atmos~\cite{Robinson_Atmos}, encoded with AC-4 Immersive Stereo (IMS)~\cite{IMS_whitepaper}, DD+JOC~\cite{purnhagen2016immersive}, and AC-4 A-JOC~\cite{ac4_IEEE}, rendered to binaural after decoding. Dolby Atmos binaural renditions served as the unencoded reference. Additionally, tests included the 3GPP IVAS codec~\cite{IVAS_AES} for ambisonics signals, with references and decoded signals rendered to binaural~\cite{IVASgithub}. We collected a total of 67,505 subjective scores.

In addition, we also augmented the training dataset by swapping the left and right channels of all audio signals while preserving the quality labels, as shown to be effective~\cite{StereoInSE-NET}.

\subsection{Test sets}
\label{subsection:testset}

Like most prior studies in RF quality modeling, we did not use no-reference subjective tests for evaluating our model due to the lack of reliable no-reference test sets for audio codecs. Our focus on evaluating codecs led us to use subjective MUSHRA tests for evaluation. We benchmarked RF-GML’s prediction accuracy against subjective scores from Unified Speech and Audio Coding (USAC)~\cite{MPEG-USAC} verification listening tests~\cite{usac_lt},\cite{USAC}, which included 24 excerpts coded with USAC, HE-AAC, and AMR-WB+ across bitrates from 8 kb/s mono to 96 kb/s stereo. The tests covered mono at low bitrates and stereo at both low and high bitrates, with 66 listeners for the mono test, 44 for the stereo low-bitrate test, and 28 for the stereo high-bitrate test. We included the mono test to evaluate RF-GML’s accuracy, as it was only trained with coded stereo and binaural tests.

Due to the unavailability of relevant binaural listening tests, we used two internal MUSHRA tests: Binaural Test-1 and Test-2, with 9 and 11 subjects, respectively. Test-1 had 11 excerpts coded with two DD+JOC variants at 448 kb/s and two IMS variants at 256 kb/s, while Test-2 included 12 excerpts coded with IMS at 64, 112, and two 256 kb/s variants. None of these excerpts were part of the training set. We computed 95\% confidence intervals using the t-distribution for all test sets.

\section{Experiments and Results}
\label{section:Results}

\subsection{Model architecture}
\label{subsection:model_arch}

Like GML, RF-GML uses an inception-based convolutional model for stereo~\cite{StereoInSE-NET} as the backbone, with architectural details provided in Fig. 3 and Table 1 of~\cite{StereoInSE-NET}, and design motivation discussed in~\cite{InSE-NET}. In short, the model combines four modified Inception blocks (In-A, In-A, In-B, In-C) adapted for audio, interspersed with three Squeeze-and-Excitation (SE) blocks, followed by three fully connected layers. The first two Inception blocks share the same kernel shapes, so both are named In-A. Unlike GML, RF-GML processes only Gammatone spectrograms for the degraded left ($L$), right ($R$), mid ($M = (L+R)/2$), and side ($S = (L-R)/2$) signals. The output stage, similar to GML, is adapted to provide two parameters describing the distribution of MUSHRA scores, as outlined in Section~\ref{section:RF-GML}. As with GML, individual listener scores are used for training rather than mean subjective scores. Our proposed RF-GML has 15.25M parameters. Since GML likely learned useful weights for predicting coded audio quality, we propose transferring the weights of degraded input channels from the first In-A block in GML and retraining with listening tests. The benefits of the transfer learning approach are demonstrated in Section~\ref{subsection:benchmark}.

\subsection{Training configuration}
\label{subsection:training_config}

The training dataset was normalized and randomly split into 80\% for training and 20\% for validation. To fully utilize the listening scores, we applied 5-fold cross-validation. The model was implemented in PyTorch and trained for 10 epochs per fold (50 epochs in total) on an Nvidia V100 GPU using the Adam optimizer. We used the optimal kernel sizes from~\cite{InSE-NET},\cite{StereoInSE-NET}, a learning rate of $10^{-4}$, and a batch size of 8, following the initialization scheme described in Sections~\ref{section:RF-GML} and~\ref{subsection:benchmark}. We also used CutMix as a data augmentation strategy, which randomly cuts and pastes spectrogram regions to create new spectrograms with associated quality scores $\tilde{y}$:
\begin{equation} 
\tilde{y} = \lambda y_{A} + (1-\lambda) y_{B}, 
\end{equation}
which is a weighted linear combination of the per-listener quality scores ($y_{A}$ and $y_{B}$) of the two involved spectrograms~\cite{gml}. The CutMix operation is performed on the fly during training, with the ratio of the cut-out area to the remaining area of the spectrogram determined by the hyperparameter $\lambda \sim \mathbf{B}(\alpha,\alpha)$ (a Beta distribution). The optimal $\alpha$ value for RF-GML was found to be 0.7, as in GML.

The model was trained using NLL loss and evaluated on NLL loss, Pearson's correlation coefficient ($R_p$), and Spearman's correlation coefficient ($R_s$). The correlation coefficients $R_p$ and $R_s$ assess the linear and monotonic relationships between variables, respectively, with $R_s$ being particularly suited for measuring rank preservation.

\subsection{RF-GML benchmarking and evaluation}
\label{subsection:benchmark}

\begin{table*}[t]
	\setlength\tabcolsep{5pt}
	\centering
	\caption{The performance of RF-GML on USAC verification and binaural listening tests is summarized below. The table shows the correlation coefficients ($R_p$ and $R_s$)$\uparrow$ between predicted and subjective mean MUSHRA scores, as well as the mean of unencoded audio (MU)$\uparrow$ for FR models (top 2 rows), RF models across all content types (middle 5 rows), and RF models across speech only (bottom 3 rows).}
	
	\vspace{-0.2cm}
	\begin{center}
		\scriptsize{
			\begin{tabular}{|l||c|c|r|c|c|r|c|c|r|c|c|r|c|c|r|}
				\hline
				\multirow{2}{*}{\backslashbox{\textbf{Models}}{\textbf{Metric}}}   & \multicolumn{3}{c|}{\textbf{Mono Bitrates}} & \multicolumn{3}{c|}{\textbf{Stereo Low Bitrates}}   & \multicolumn{3}{c|}{\textbf{Stereo High Bitrates}}      & \multicolumn{3}{c|}{\textbf{Binaural Test-1}}   & \multicolumn{3}{c|}{\textbf{Binaural Test-2}} \\ \cline{2-16} 
				& $\mathbf{R_p}$   & $\mathbf{R_s}$ & \multicolumn{1}{c|}{$\mathbf{MU}$} & $\mathbf{R_p}$   & $\mathbf{R_s}$ & \multicolumn{1}{c|}{$\mathbf{MU}$} & $\mathbf{R_p}$   & $\mathbf{R_s}$ & \multicolumn{1}{c|}{$\mathbf{MU}$} & $\mathbf{R_p}$   & $\mathbf{R_s}$ & \multicolumn{1}{c|}{$\mathbf{MU}$} & $\mathbf{R_p}$   & $\mathbf{R_s}$ & \multicolumn{1}{c|}{$\mathbf{MU}$}\\ \hline\hline
	    \rowcolor{gray!20} 
    \textbf{FR models} \\ \hline			
    \textbf{ViSQOL-v3}      & 0.81  & 0.84  & 94.64  & 0.77  & 0.78  & 94.64   & 0.82  & 0.90  & 94.64   & 0.90  & 0.93  & 94.64   & 0.96  & 0.85  & 94.64 \\ \hline
				\textbf{GML}  & \textbf{0.88}  & \textbf{0.88}  & \textbf{100} & \textbf{0.89}  & \textbf{0.86}  & \textbf{100}  & \textbf{0.92}  & \textbf{0.94}  & \textbf{100}  & \textbf{0.98}  & \textbf{0.95}  & \textbf{100}  & \textbf{0.98}  & \textbf{0.92}  & \textbf{100}\\ \hline
    \hline\hline
    \rowcolor{gray!20} 
    \textbf{RF models} \\ \hline
    \textbf{SESQA}              & 0.26  & 0.26  & 64.17 & 0.28  & 0.31  & 64.23  & 0.22  & 0.22  & 64.23  & 0.32  & 0.33  & 54.06  & 0.14  & 0.17  & 45.36 \\ \hline
				\textbf{RF-GML (def)}       & \textbf{0.82}  & \textbf{0.83}  & 82.14 & 0.78  & \textbf{0.77}  & 84.54  & 0.78  & 0.65  & 84.54  & \textbf{0.86}  & 0.68  & 82.43  & \textbf{0.97}  & \textbf{0.81}  & 96.25 \\ \hline
     \rowcolor{mycolor}     
				\textbf{RF-GML (deg)}       & 0.78  & 0.76  & \textbf{89.56} & \textbf{0.81}  & 0.75  & \textbf{93.78}  & \textbf{0.86}  & \textbf{0.81}  & \textbf{93.78}  & 0.79  & 0.73  & \textbf{88.82}  & 0.90  & 0.76  & \textbf{99.38} \\ \hline
				\textbf{RF-GML (degF)}    & 0.76  & 0.75  & 78.79 & 0.79  & 0.75  & 81.57  & 0.85  & 0.79  & 81.57  & 0.84  & 0.71  & 79.67  & 0.95  & 0.79  & 84.35 \\ \hline
				\textbf{RF-GML (all)}       & 0.73  & 0.72  & 71.04 & 0.76  & 0.71  & 71.38  & 0.81  & 0.70  & 71.38  & \textbf{0.86}  & \textbf{0.79}  & 70.58  & 0.93  & 0.76  & 68.21 \\ \hline
    \hline\hline
    \rowcolor{gray!20} 
    \textbf{Performance on speech} \\ \hline
               \textbf{SESQA}              & \textbf{0.82}  & \textbf{0.83}  & 82.29 & \textbf{0.78}  & \textbf{0.80}  & 81.48  & 0.69  & 0.59  & 81.48  & n.a.  & n.a. & n.a.   & n.a.  & n.a. & n.a.  \\ \hline
                \textbf{RF-GML (def)}       & 0.79  & 0.78  & 90.95 & 0.71  & 0.63  & 92.18  & 0.79  & 0.71  & 92.18 & n.a.  & n.a. & n.a.   & n.a.  & n.a. & n.a.   \\ \hline
                \rowcolor{mycolor}
				\textbf{RF-GML (deg)}       & 0.75  & 0.74  & \textbf{93.21} & 0.73  & 0.65  & \textbf{93.02}  & \textbf{0.84}  & \textbf{0.79}  & \textbf{93.02} & n.a.  & n.a. & n.a.   & n.a.  & n.a. & n.a.    \\ \hline
		\end{tabular}}
	\end{center}


\end{table*}

We benchmark RF-GML against two FR metrics: ViSQOL-v3 (audio mode)~\cite{ViSQOLgithub} and GML (also trained with CutMix)~\cite{gml}. ViSQOL-v3 is included due to its strong correlation with subjective scores in audio codec evaluations~\cite{Fraunhofer}. GML is included for comparison as RF-GML is derived from it and trained on the same dataset. As noted in Section~\ref{section:Introduction}, there is no RF model for general audio at a 48 kHz sample rate. Research in~\cite{no-ref-audio-lossy} focuses on mono music coding at 48 kHz only; however, the shared code is not well-documented for evaluating an audio signal, evaluation is restricted to 10s excerpts, and it requires running 15 model variants. Hence, we use SESQA, a state-of-the-art RF model for mono speech at 48 kHz.

We evaluated the accuracy of mean quality score predictions using $R_s$ for monotonicity and $R_p$ for linearity, with higher values indicating better performance. We also calculated the mean quality score of the unencoded audio (MU), where a higher value signifies better performance. We did not assess CI prediction accuracy because we’re evaluating our model against MUSHRA tests. There is no prior research on how MUSHRA ratings would change if systems were re-evaluated without an open reference, but we speculate (based on~\cite{schepker2020acoustic}) that system rankings might be preserved, although the task’s increased difficulty could lead to higher CIs. Hence, we leave the evaluation of CI prediction as future work after constructing corresponding subjective tests without an open reference~\cite{MUSHRA_Noref}.

The results are shown in Table I, with the best-performing models highlighted in bold. For SESQA and ViSQOL, MU was computed by multiplying their MOS predictions by 20. Note that ViSQOL downmixes two-channel (stereo and binaural) signals to a mono mid-signal for quality prediction. Similarly, since SESQA is designed for mono, we also downmix stereo and binaural signals to a mono mid-signal for assessment. GML and RF-GML can evaluate stereo and binaural, so for mono, we assess a dual-mono signal (with $L = R$) with GML and RF-GML.

We compared four variants of RF-GML: (1) RF-GML (def) was trained from scratch using the default PyTorch initializer~\cite{He_2015}, i.e., without transfer learning. (2) In RF-GML (deg), the weights of degraded input channels in the first inception (In-A) block were initialized with corresponding degraded input channels from a pre-trained GML. (3) RF-GML (degF) is identical to RF-GML (deg), but its initialized weights remained frozen during training. (4) In RF-GML (all), all four inception blocks were initialized with weights from the GML.

\begin{figure}[t]
	\centering
	\includegraphics[width=0.85\linewidth]{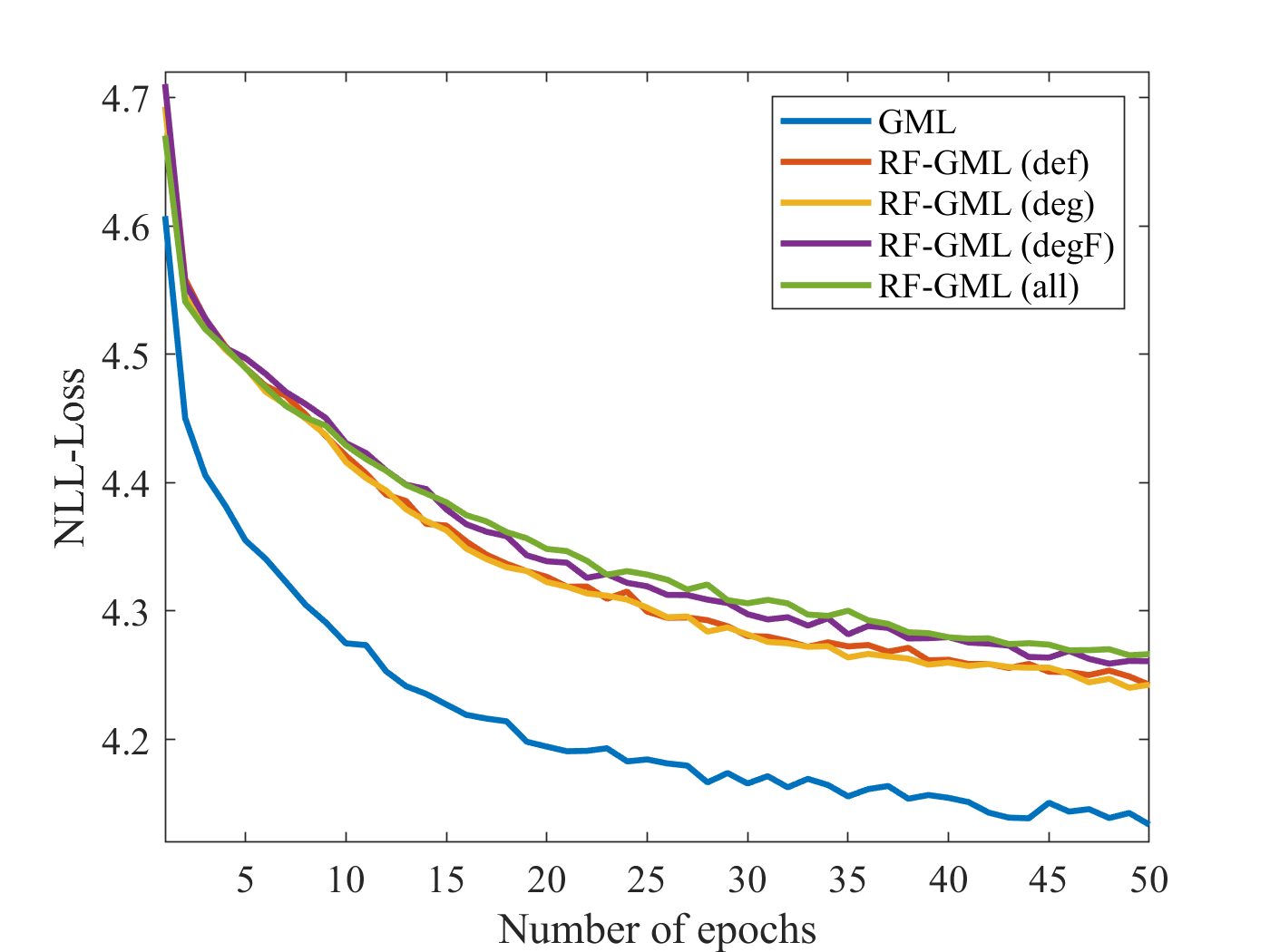} 
	\caption{NLL training losses of GML and RF-GML flavors.}
	\label{fig:TrainingLosses}
\end{figure}

Figure~\ref{fig:TrainingLosses} shows that GML has the lowest NLL loss and is expected to have the highest prediction accuracy. Table 1 further confirms its superior accuracy over ViSQOL~\cite{gml}. While RF-GML (def) and RF-GML (deg) have similar training losses, Table 1 indicates that RF-GML (deg) performs best across all test sets and metrics, often achieving top-1 scores and reliably rating unencoded audio closer to 100. We speculate this is due to knowledge transfer from GML, which was trained on many reference-reference signal pairs. Although RF-GML (deg) may have lower correlation scores at times compared to RF-GML (def), considering all the test sets and metrics, it outperforms RF-GML (def). RF-GML (degF) shows decreased prediction accuracy when initialized like RF-GML (def) but frozen, underscoring the need for both initialization and learning from the first inception (In-A) block. Simply initializing all inception block weights from GML further degrades unencoded reference ratings. Due to slightly poorer NLL losses, both RF-GML (degF) and RF-GML (all) are excluded from further analysis. On average, considering all five listening test sets together (evaluating 953 signals), RF-GML (deg) achieves $R_p$, $R_s$, and MU of 0.83, 0.83, and 89.56, respectively, compared to the second-best RF-GML (def), which achieves 0.83, 0.81, and 82.14, respectively. Table 1 shows that, for the mono and stereo tests, these two RF-GML models achieve a correlation with subjective ratings comparable to ViSQOL.

SESQA, trained specifically for speech, performs poorly on general audio, so we also benchmarked its accuracy against RF-GML (deg) and RF-GML (def) models using all eight speech excerpts from each listening test. Since the binaural tests lack speech signals, no results are available in those cases. SESQA shows improved $R_p$ and $R_s$ scores on low-bitrate mono coding and, interestingly, on the low-bitrate stereo test. However, RF-GML outperforms SESQA in the high-bitrate stereo test, and SESQA still struggles to rate unencoded speech closer to 100.

Next, we encoded all 24 stereo excerpts from the test set at various bitrates using the HE-AAC codec family commonly applied in practice, and demonstrated (Figure 2) that the mean RF-GML quality score scales well with bitrate. However, we notice that the average quality predicted with RF-GML (deg) at 96 kb/s HE-AAC v1 is slightly higher than that of AAC at 192 kb/s. A possible reason could be that HE-AAC v1 at 96 kb/s has a higher overall bandwidth due to the use of spectral band replication~\cite{HE-AAC}. Since we would expect the quality of both 192 kb/s AAC and 96 kb/s HE-AAC to fall in the excellent quality range, this would have been a difficult task even for human evaluation in a reference-free quality assessment. On the other hand, RF-GML (def) predictions saturate around 80 on this data, with less quality separation across bitrates and a slight quality increase when dropping from 32 to 20 kb/s. This observation, along with strong performance in predicting subjective quality (as shown in Table I), makes RF-GML (deg) more suitable for quality monitoring in adaptive streaming clients.

\begin{figure}[t]
  \centering
  \includegraphics[width=0.83\linewidth]{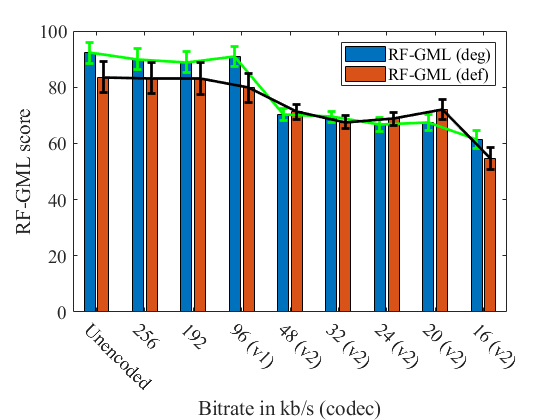} 
  \caption{Scaling of RF-GML scores with bitrates: 24 stereo excerpts~\cite{usac_lt} coded with HE-AAC (v1 or v2) and AAC.}
  \label{fig:quality_scaling}
\end{figure}

\begin{figure}[t]
	\centering
	\includegraphics[width=0.8\linewidth]{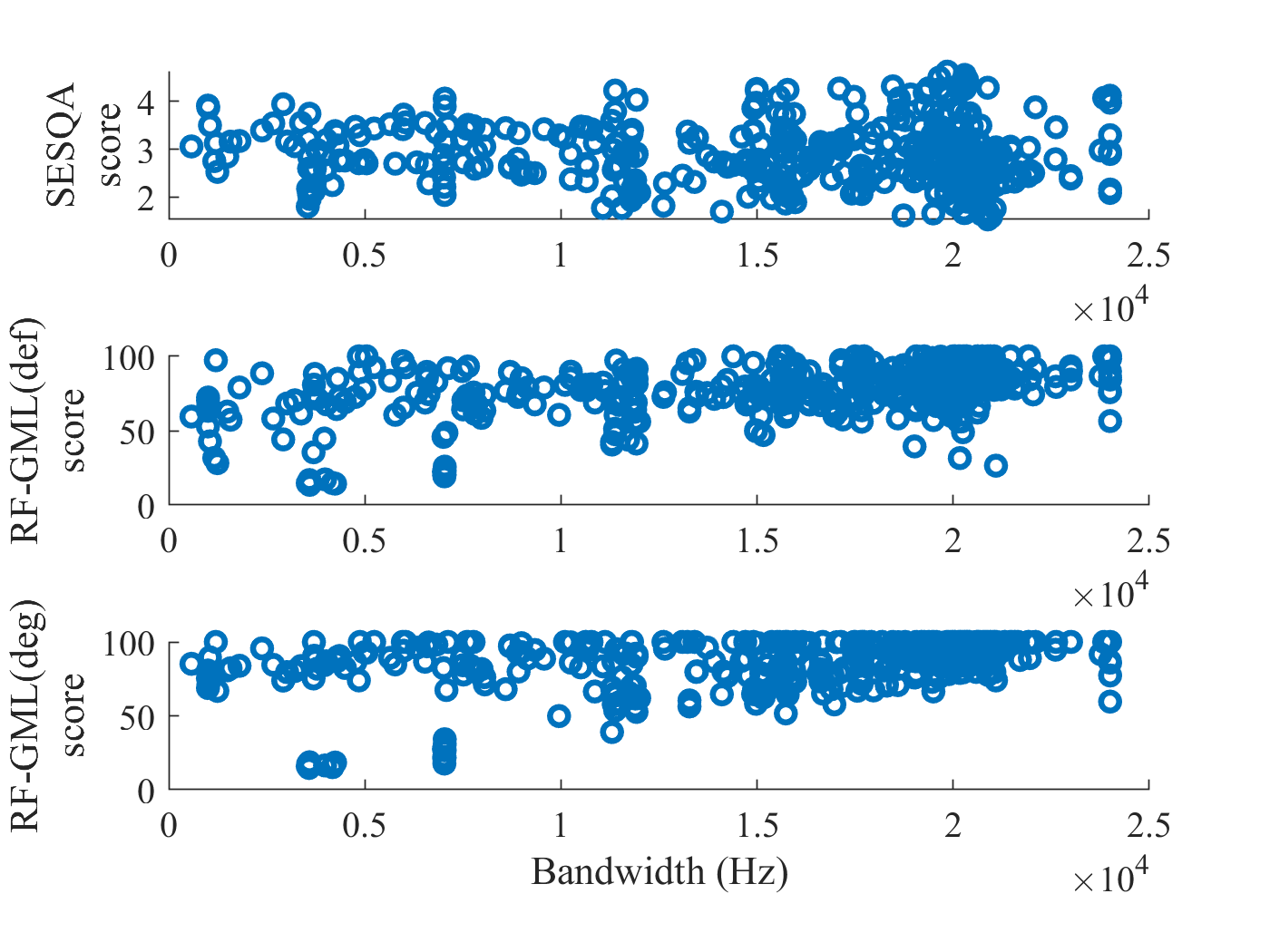} 
	\caption{Correlation of predicted quality scores versus audio bandwidth: -0.08 (SESQA), 0.51 (RF-GML (def)), 0.44 (RF-GML (deg)), with RF-GML (deg) rating unencoded audio closer to 100.}
	\label{fig:scatter_plot}
\end{figure}

Finally, we evaluated the model’s performance in rating unencoded content by running RF-GML and SESQA on 511 excerpts, including synthetic test tones and signals, and plotted predicted quality scores versus signal bandwidth in scatter plots (Figure~\ref{fig:scatter_plot}). This content, curated internally over the years for testing the engineering implementation of codecs, was not fully known to the authors before the experiments. Figure~\ref{fig:scatter_plot} shows almost no correlation (-0.08) between bandwidth and quality scores for SESQA, while RF-GML (deg) and RF-GML (def) show slight correlations of 0.44 and 0.51, respectively. It is evident that SESQA is unsuitable for rating unencoded content. A higher correlation with bandwidth is expected with the RF-GML variants, as they are trained to evaluate coding artifacts, with audio bandwidth being a typical codec tuning parameter across bitrates (or quality levels). However, note that a better correlation with RF-GML (def) may not necessarily indicate better performance, as unencoded speech signals may also have lower bandwidth (e.g., 16 kHz), though ideally, the predicted quality should be close to 100. The scatter plot further shows that RF-GML (deg) rates unencoded audio closer to 100 than RF-GML (def). The five lowest-rated signals for RF-GML are the 3.5 kHz and 4.5 kHz low-pass filtered versions, while SESQA scored full-bandwidth unencoded music signals the lowest. This is expected for RF-GML, which was trained on MUSHRA tests with low-pass filtered anchors, and for SESQA, which was not trained with music signals. Both SESQA and RF-GML (deg) rated the five lowest bandwidth signals (1 kHz sine tones and an $\approx$500 Hz tone from a gong instrument) similarly, assigning high-quality scores. These results demonstrate that RF-GML (deg) consistently rates unencoded audio closer to 100. 

\section{Conclusions}
\label{section:ConclusionDiscussion}

This paper presents RF-GML, a novel reference-free audio quality metric capable of evaluating coded mono, stereo, and binaural audio at 48 kHz. Extensive benchmarking demonstrates RF-GML’s superior performance. The transfer learning approach, which leverages knowledge from a pre-trained full-reference GML, has proven effective in enhancing RF-GML’s accuracy. By initializing the weights of degraded input channels with corresponding weights from the full-reference model, RF-GML gains a better understanding of reference signals, leading to more precise quality predictions. The proposed metric offers several advantages: it is versatile and capable of evaluating a wide range of audio content and coding scenarios. While RF-GML can predict confidence intervals, this aspect has not been evaluated in this paper and will be explored in future work, following the construction of reference-free subjective tests.

\bibliographystyle{IEEEtran}
\bibliography{IEEEabrv,IEEEexample}

\begin{thebibliography}{10}
\providecommand{\url}[1]{#1}
\csname url@samestyle\endcsname
\providecommand{\newblock}{\relax}
\providecommand{\bibinfo}[2]{#2}
\providecommand{\BIBentrySTDinterwordspacing}{\spaceskip=0pt\relax}
\providecommand{\BIBentryALTinterwordstretchfactor}{4}
\providecommand{\BIBentryALTinterwordspacing}{\spaceskip=\fontdimen2\font plus
\BIBentryALTinterwordstretchfactor\fontdimen3\font minus \fontdimen4\font\relax}
\providecommand{\BIBforeignlanguage}[2]{{%
\expandafter\ifx\csname l@#1\endcsname\relax
\typeout{** WARNING: IEEEtran.bst: No hyphenation pattern has been}%
\typeout{** loaded for the language `#1'. Using the pattern for}%
\typeout{** the default language instead.}%
\else
\language=\csname l@#1\endcsname
\fi
#2}}
\providecommand{\BIBdecl}{\relax}
\BIBdecl

\bibitem{peaq}
T.~Thiede \emph{et~al.}, ``{PEAQ}---the {ITU} standard for objective measurement of perceived audio quality,'' \emph{Journal of the Audio Engineering Society}, vol.~48, pp. 3--29, 01 2000.

\bibitem{ViSQOLAudioIEEE}
C.~Sloan \emph{et~al.}, ``Objective assessment of perceptual audio quality using {ViSQOLAudio},'' \emph{IEEE Transactions on Broadcasting}, vol.~63, no.~4, pp. 693--705, 2017.

\bibitem{polqa}
J.~Beerends \emph{et~al.}, ``Perceptual objective listening quality assessment {POLQA}, the third generation {ITU-T} standard for end-to-end speech quality measurement part {I}-temporal alignment,'' \emph{AES: Journal of the Audio Engineering Society}, vol.~61, pp. 366--384, 06 2013.

\bibitem{non-intru_ms}
A.~R. Avila \emph{et~al.}, ``Non-intrusive speech quality assessment using neural networks,'' in \emph{ICASSP 2019 - 2019 IEEE International Conference on Acoustics, Speech and Signal Processing (ICASSP)}, 2019, pp. 631--635.

\bibitem{Fu2018QualityNetAE}
S.~wei Fu \emph{et~al.}, ``{Quality-Net}: An end-to-end non-intrusive speech quality assessment model based on {BLSTM},'' in \emph{Proc. Interspeech 2018}, 2018, pp. 1873--1877.

\bibitem{Wawenets}
A.~A. Catellier and S.~D. Voran, ``Wawenets: A no-reference convolutional waveform-based approach to estimating narrowband and wideband speech quality,'' in \emph{ICASSP 2020 - 2020 IEEE International Conference on Acoustics, Speech and Signal Processing (ICASSP)}, 2020, pp. 331--335.

\bibitem{NISQA}
G.~Mittag and S.~Möller, ``Non-intrusive speech quality assessment for super-wideband speech communication networks,'' in \emph{ICASSP 2019 - 2019 IEEE International Conference on Acoustics, Speech and Signal Processing (ICASSP)}, 2019, pp. 7125--7129.

\bibitem{SESQA}
J.~Serrà \emph{et~al.}, ``{SESQA}: Semi-supervised learning for speech quality assessment,'' in \emph{ICASSP 2021 - 2021 IEEE International Conference on Acoustics, Speech and Signal Processing (ICASSP)}, 2021, pp. 381--385.

\bibitem{noresqa}
P.~Manocha \emph{et~al.}, ``{NORESQA}: A framework for speech quality assessment using non-matching references,'' in \emph{Thirty-Fifth Conference on Neural Information Processing Systems}, 2021.

\bibitem{nomad}
A.~Ragano \emph{et~al.}, ``{NOMAD}: Unsupervised learning of perceptual embeddings for speech enhancement and non-matching reference audio quality assessment,'' in \emph{ICASSP 2024 - 2024 IEEE International Conference on Acoustics, Speech and Signal Processing (ICASSP)}, 2024, pp. 1011--1015.

\bibitem{no-ref-pesq-dnn}
Z.~Xu \emph{et~al.}, ``Coded speech quality measurement by a non-intrusive {PESQ-DNN},'' \emph{IEEE/ACM Transactions on Audio, Speech, and Language Processing}, vol.~31, pp. 3404--3417, 2023.

\bibitem{Incept2}
C.~{Szegedy} \emph{et~al.}, ``Rethinking the inception architecture for computer vision,'' in \emph{2016 IEEE Conference on Computer Vision and Pattern Recognition (CVPR)}, 2016, pp. 2818--2826.

\bibitem{InSE-NET}
G.~{Jiang} \emph{et~al.}, ``{InSE-NET}: A perceptually coded audio quality model based on cnn,'' in \emph{151st AES Convention}, October 2021.

\bibitem{StereoInSE-NET}
A.~{Biswas} and G.~{Jiang}, ``Stereo {InSE-NET}: Stereo audio quality predictor transfer learned from mono inse-net,'' in \emph{153rd AES Convention}, October 2022.

\bibitem{gml}
G.~{Jiang} \emph{et~al.}, ``Generative machine listener,'' in \emph{155th AES Convention}, October 2023.

\bibitem{Mumtaz}
D.~Mumtaz \emph{et~al.}, ``Nonintrusive perceptual audio quality assessment for user-generated content using deep learning,'' \emph{IEEE Transactions on Industrial Informatics}, vol.~18, no.~11, pp. 7780--7789, 2022.

\bibitem{no-ref-audio-lossy}
A.~Kasperuk and S.~K. Zieliński, ``Non-intrusive method for audio quality assessment of lossy-compressed music recordings using convolutional neural networks,'' \emph{International Journal of Electronics and Telecommunications}, vol.~70, no.~2, pp. 331--339, 2024.

\bibitem{ViSQOLgithub}
Google. (2022) {ViSQOL}. \url {https://github.com/google/visqol/}.

\bibitem{MUSHRA}
``Method for the subjective assessment of intermediate quality level of audio systems,'' International Telecommunication Union, Standard Recommendation ITU-R BS.1534-3, 2015.

\bibitem{usac_lt}
``{USAC} verification test report,'' International Organisation for Standardisation, Torino, Italy, Tech. Rep. ISO/IEC JTC1/SC29/WG1 MPEG2011/N12232, 2011.

\bibitem{faridee22_MOSdistr}
A.~Z.~M. Faridee and H.~Gamper, ``{Predicting label distribution improves non-intrusive speech quality estimation},'' in \emph{Proc. Interspeech 2022}, 2022, pp. 406--410.

\bibitem{AAC}
M.~Bosi \emph{et~al.}, ``{ISO/IEC} {MPEG-2} advanced audio coding,'' \emph{Journal of the Audio Engineering Society}, vol.~45, no.~10, pp. 789--814, October 1997.

\bibitem{HE-AAC}
A.~C. den Brinker \emph{et~al.}, ``An overview of the coding standard {MPEG-4} audio amendments 1 and 2: {HE-AAC}, {SSC}, and {HE-AAC} v2,'' \emph{EURASIP Journal on Audio, Speech, and Music Processing}, vol. 2009, no.~1, 2009.

\bibitem{ac4_IEEE}
J.~Riedmiller \emph{et~al.}, ``Delivering scalable audio experiences using {AC-4},'' \emph{IEEE Transactions on Broadcasting}, vol.~63, no.~1, pp. 179--201, 2017.

\bibitem{Robinson_Atmos}
C.~Q. Robinson \emph{et~al.}, ``Scalable format and tools to extend the possibilities of cinema audio,'' \emph{SMPTE Motion Imaging Journal}, vol. 121, no.~8, pp. 63--69, 2012.

\bibitem{IMS_whitepaper}
\BIBentryALTinterwordspacing
{Dolby AC-4 Whitepaper}. (2021, February) Dolby {AC-4}: Audio delivery for next-generation entertainment services. [Online]. Available: \url{https://professional.dolby.com/siteassets/technologies/dolbt_atmos_ac-4_whitepaper.pdf}
\BIBentrySTDinterwordspacing

\bibitem{purnhagen2016immersive}
H.~{Purnhagen} \emph{et~al.}, ``Immersive audio delivery using joint object coding,'' in \emph{140th AES Convention}, May 2016.

\bibitem{IVAS_AES}
M.~Multrus \emph{et~al.}, ``Immersive voice and audio services ({IVAS}) codec - the new {3GPP} standard for immersive communication,'' in \emph{157th AES Convention}, October 2024.

\bibitem{IVASgithub}
3GPP. (2022) {IVAS Codec Public Collaboration}. \url {https://forge.3gpp.org/rep/ivas-codec-pc/ivas-codec}.

\bibitem{MPEG-USAC}
S.~{Quackenbush}, ``{MPEG} unified speech and audio coding,'' \emph{IEEE MultiMedia}, vol.~20, no.~2, pp. 72--78, 2013.

\bibitem{USAC}
S.~{Quackenbush} and R.~{Lefevbre}, ``Performance of {MPEG} unified speech and audio coding,'' in \emph{131st AES Convention}, October 2011.

\bibitem{Fraunhofer}
P.~M. {Delgado} and J.~{Herre}, ``Can we still use {PEAQ}? a performance analysis of the {ITU} standard for the objective assessment of perceived audio quality,'' in \emph{2020 Twelfth International Conference on Quality of Multimedia Experience (QoMEX)}, 2020, pp. 1--6.

\bibitem{schepker2020acoustic}
H.~Schepker \emph{et~al.}, ``Acoustic transparency in hearables—perceptual sound quality evaluations,'' \emph{Journal of Audio Engineering Society}, vol.~68, pp. 495--507, july 2020.

\bibitem{MUSHRA_Noref}
``Method for the subjective quality assessment of audible differences of sound systems using multiple stimuli without a given reference,'' International Telecommunication Union, Standard Recommendation ITU-R BS.2132-0, 2019.

\bibitem{He_2015}
K.~He \emph{et~al.}, ``Delving deep into rectifiers: Surpassing human-level performance on {ImageNet} classification,'' in \emph{2015 IEEE International Conference on Computer Vision (ICCV)}, 2015, pp. 1026--1034.

\end{thebibliography}

\end{document}